\documentclass[a4paper,print,showpacs,prb,twocolumn]{revtex4}
\usepackage{bm}
\usepackage{amsmath}
\usepackage{graphicx}
\usepackage{amsfonts}
\usepackage{amssymb}

\input{tcilatex}
\begin{document}

\title{Anisotropic transport in quantum wires embedded in (110) plane}
\author{Fang Cheng$^{1,2,3}$}
\email{fcheng@semi.ac.cn}
\author{Kai Chang$^{1}$}
\email{kchang@semi.ac.cn} \affiliation{$^{1}$SKLSM, Institute of
Semiconductors, Chinese Academy of Sciences, P. O. Box 912, Beijing
100083, China} \affiliation{$^{2}$Department of Physics and
Electronic Science, Changsha University of Science and Technology,
Changsha 410076, China} \affiliation{$^{3}$KLLDQSQC, Hunan Normal
University, Changsha 410081, China}
\date{\today }

\begin{abstract}
We investigate theoretically the effects of the Coulomb interaction
and spin-orbit interactions (SOIs) on the anisotropic transport
property of semiconductor quantum wires embedded in (110) plane. The
anisotropy of the dc conductivity can be enhanced significantly by
the Coulomb interaction for infinite-long quantum wires. But it is
smeared out in quantum wires with finite length, while the ac
conductivity still shows anisotropic behavior, from which one can
detect and distinguish the strengths of the Rashba SOI and
Dresselhaus SOI.
\end{abstract}

 \maketitle

All-electrical manipulation of spin degree of freedom is one of the
central issues and the ultimate goal of spintronics field. The
spin-orbit interaction (SOI) provides us an efficient way to control
electron spin electrically and therefore has attracted tremendous
interest from the view of point both the potential application in
all-electrical controlled spintronic devices and fundamental
physics.$^{1}$ The SOI is manifest of the
relativistic effect and caused by the broken of spatial inversion symmetry.$%
^{2}$ The spatial inversion symmetry can be broken by the structure
inversion symmstry (SIA) and bulk crystal inversion symmetry, named
Rashba SOI (RSOI) and Dresselhaus SOI (DSOI),
respectively.$^{3\text{,}4}$ In thin quantum wells, the strength of
the DSOI is comparable to that of the RSOI since the strength of the
DSOI depends significantly on the thickness of quantum wells. The
interplay between the RSOI and the DSOI leads to interesting
phenomena, e.g., the anisotropic photogalvanic effect$^{5}$, the
persistent spin helix$^{6}$, and anisotropic transport property of
quantum wire$^{7,8}$.

Changing the crystallographic planes can have a significant effect
on the interplay between RSOI and DSOI, which is a consequence of
the fact that the DSOI depends sensitively on the crystallographic
planes electrons are moving in. It is highly desirable to study how
the crystallographic plane affects the transport property, which is
interesting and important for potential applications of
semiconductor spintronic devices. Very recently, the anisotropic
behavior of transport property in semiconductor quantum wire was
proposed to detect the relative strength between the RSOI and DSOI
in a quasi-one-dimensional (Q1D) semiconductor quantum wire
system.$^{7}$ But the effect of the Coulomb interaction on the
transport property is not addressed. Since the Coulomb interaction
becomes very important for Q1D systems where electrons are strongly
correlated, the conventional Fermi liquid theory breaks down. We
will address this issue in this Letter based on the Luttinger liquid
(LL) theory$^{9}$. The LL is of fundamental importance because it is
one of very few strongly correlated non-Fermi liquid systems that
can be solved analytically. The RSOI would lead to the mixing
between the spin and charge excitations.$^{10-13}$ The LL\ behavior
was demonstrated experimentally in many Q1D system, e.g., narrow
quantum
wire formed in semiconductor heterostructures$^{14}$, carbon nanotube$^{15}$%
, graphene nanoribbon$^{16}$, as well as the edge states of the
fractional Quantum Hall liquid$^{17}$.

In this Letter, we investigate theoretically the Coulomb interaction
on the anisotropic transport properties in semiconductor quantum
wires at (110) crystallographic plane in the presence of the RSOI
and DSOI. We find that, in contrary to the non-interacting electron
gas, the anisotropy of the dc transport property is smeared out for
a quantum wire with finite length. The ac conductivity still depends
sensitively on the crystallographic direction, which provides us a
possible way to detect the strengths of the RSOI and DSOI in Q1D
semiconductor quantum wire.

\begin{figure}[b]
\includegraphics[width=2.6in]{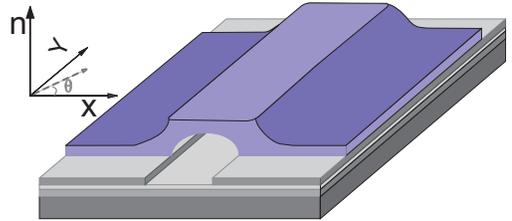}
\caption{(color online) Schematic diagram of semiconductor quantum
wire which can be fabricated along the crystallographic direction
$\protect\theta $ with respect to $[100]$ axis in the
crystallographic planes $(110)$.} \label{fig:fig1}
\end{figure}
For a clean Q1D quantum wire embedded in (110) plane (see Fig. 1),
the Hamiltonian of the noninteracting electrons reads$^{8}$
\begin{equation}
H_{0}=\frac{\hbar ^{2}k_{x}^{2}}{2m^{\ast }}+V(r)-\alpha \sigma _{y}k_{x}-%
\frac{1}{2}\beta sin\theta \sigma _{z}k_{x},  \label{myeq1}
\end{equation}%
where $\theta $ is the angle between the orientation of the quantum
wire and
the [100] axis, $m^{\ast }$ is the electron effective mass, $\sigma _{i}$ ($%
i $=$x,y,z$) are the Pauli matrices, $\alpha $ and $\beta $ are the
strengths of the RSOI and DSOI, respectively. Assuming $\delta v\ll v_{F}$ ($%
v_{F}$ is the bare Fermi velocity of right and left moving
noninteracting electrons), the linearized noninteracting electron
Hamiltonian of the quantum wire with both RSOI and DSOI is given
by$^{10,13}$ $H_{0}=-i\hbar \int \sum_{\gamma ,s}v_{\gamma }^{s}\psi
_{\gamma s}^{+}\partial _{x}\psi _{\gamma s}dx$, where the operators
$\psi _{\gamma s}$ $(\gamma
=-1(L),1(R);s=-1(\downarrow ),1(\uparrow ))$ annihilate spin-down $%
(\downarrow )$ or spin-up $(\uparrow ) $ electrons near the left (L)
and right $(R)$ Fermi points, respectively. $v_{\gamma }^{s}=\gamma
v_{F}-s\delta v$ are the four different Fermi velocities, where $\delta v$=$%
\sqrt{\alpha ^{2}+\beta ^{2}sin^{2}\theta /4}/\hbar $. Note that the
RSOI and DSOI split the spin subbands and make the electron Fermi
velocities become different for different directions of motion. The
total Hamiltonian of the system is $H=H_{0}+H_{int}$, where
$H_{int}\!\!=\!\!\frac{1}{2}\int \int dxdy\psi _{\gamma s}^{\dagger
}(x)\psi _{\gamma ^{^{\prime }}\!s^{^{\prime }}}^{\dagger
}(y)V\!\!_{\gamma \gamma ^{^{\prime }}}(x-y)\psi _{\gamma ^{^{\prime
}}\!s^{^{\prime }}}(y)\psi _{\gamma s}(x)$. The Umklapp scattering
process is neglected because the Fermi energy in quantum wires
formed in semiconductor heterostructure is far from the half-filled
case. And the electron-electron backscattering can be negligible for
a sufficiently long interacting region.$^{10}$ Using the
bosonization
technique$^{18}$, the Hamiltonian becomes%
\begin{eqnarray}
H &=&\frac{\hbar }{2}\int dx\Big[\frac{v_{\rho }}{K_{\rho
}}(\partial _{x}\vartheta _{\rho })^{2}+v_{\rho }K_{\rho
}\Big(\frac{\Pi _{\rho }}{\hbar
}\Big)^{2}\Big]  \notag \\
&+&\frac{\hbar }{2}\int dx\Big[\frac{v_{\sigma }}{K_{\sigma
}}(\partial
_{x}\vartheta _{\sigma })^{2}+v_{\sigma }K_{\sigma }\Big(\frac{\Pi _{\sigma }%
}{\hbar }\Big)^{2}\Big]  \notag \\
&+&\hbar \delta v\int dx\Big[\Big(\frac{\Pi _{\sigma }}{\hbar }\Big)%
(\partial _{x}\vartheta _{\rho })+\Big(\frac{\Pi _{\rho }}{\hbar }\Big)%
(\partial _{x}\vartheta _{\sigma })\Big],  \label{myeq2}
\end{eqnarray}%
where $\vartheta _{\rho }$ and $\vartheta _{\sigma }$ are the phase
fields for the charge and spin degrees of freedom, respectively, and
$\Pi _{\rho }$ and $\Pi _{\sigma }$ are the corresponding conjugate
momenta. $v_{\rho ,\sigma }$ are the propagation velocities of the
charge and spin collective modes of the decoupled model ($\delta
v=0$). In the following, we consider only pointlike density-density
interactions. $v_{\rho,\sigma }$=$v_{F}/K_{\rho ,\sigma }$ $^{18}$,
and the parameter $K_{\rho /\sigma }$ is defined as $1/K_{\rho
/\sigma }^{2}$=$1\pm g$, where $g$=2$V(q$=0)/$\hbar \pi v_{F}$ with
$V(q$=0) is the electron-electron interaction potential.

We consider an interacting Q1D quantum wire under a time-dependent
electric
field $E(x,t)$ along the wire, e.g., a microwave radiation. $H_{ac}=-\sqrt{%
\frac{2}{\pi }}e\int dxE(x,t)\vartheta _{\rho }(x,t)$ describes the
interaction between electron and radiation field in the quantum
wire.$^{19}$ The total Hamiltonian becomes $H=H_{0}+H_{int}+H_{ac}$.
Using the equation of motion and the linear response theory, we
obtain the nonlocal charge conductivity
\begin{eqnarray}
\sigma _{\rho }(x,\omega )&=&\frac{2e^{2}}{h}\Bigg[\frac{%
(u_{1}^{2}-v_{\sigma }^{2}+\delta v^{2})}{(u_{1}^{2}-u_{2}^{2})}\frac{v_{F}}{%
u_{1}}e^{i\frac{\omega }{u_{1}}|x|}
\nonumber\\&-&\frac{(u_{2}^{2}-v_{\sigma }^{2}+\delta v^{2})}{(u_{1}^{2}-u_{2}^{2})}%
\frac{v_{F}}{u_{2}}e^{i\frac{\omega }{u_{2}}|x|}\Bigg],
\label{myeq3}
\end{eqnarray}
where%
\begin{equation}
u_{1,2}^{2}\!\!=\frac{v_{\rho }^{2}+v_{\sigma }^{2}}{2}+\delta v^{2} \pm \!%
\sqrt{\Big(\frac{v_{\rho }^{2}\!-\!v_{\sigma
}^{2}}{2}\Big)^{2}\!\!+2\delta v^{2}(v_{\rho }^{2}\!+v_{\sigma
}^{2})},  \label{myeq4}
\end{equation}
where $u_{1,2}$ are the propagation velocities of coupled collective
modes that depend on the crystallographic orientation $\theta$.
Notice that the SOIs couple the spin and charge excitations in the
absence of the SOIs.

\begin{figure}
\center
\includegraphics[width=1\columnwidth]{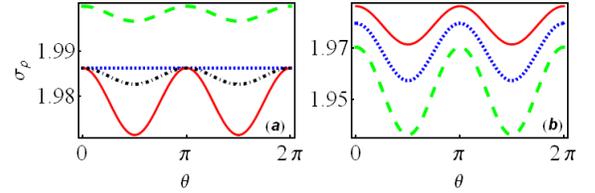} %
\caption{(color online) The dc conductivity (in units of $K_{\protect\rho %
}e^{2}/h$) as a function of the angle $\protect\theta $. (a)with fixed $g$%
=0.4 where solid red line is for $\protect\alpha $=0.2 and $\protect\beta $%
=0.4, the dot-dashed black line for $\protect\alpha $=0.2 and
$\protect\beta
$=0.2, the dashed green line for $\protect\alpha $=0 and $\protect\beta $%
=0.2, and the dotted blue line for $\protect\alpha $=0.2 and $\protect\beta $%
=0. (b)with fixed strengths of the SOI $\protect\alpha $=0.2 and $\protect%
\beta $=0.4, where the solid red line corresponds to $g$=0.4, the
dotted blue line to $g$=0.6, and the dashed green line to $g$=0.8,
respectively.}\label{fig:fig2}
\end{figure}
In dc case, Eq. (3) shows that the charge conductivity of a perfect
quantum wire with the RSOI and DSOI depend on the parameters $g$ and
$\delta v$. In the absence of the SOIs, i.e., $\alpha =\beta =0$,
the dc charge conductivity $\sigma _{\rho }=2K_{\rho }e^{2}/h$ which
is in agreement with the previous studies$^{20}$. The dc
conductivity depends sensitively on the crystallographic direction
$\theta $ of the waveguide structure, i.e., the anisotropic
transport behavior [see the solid red line, dashed green line, and
black dot-dashed in Fig. 2(a)]. The anisotropy is caused by the
interplay between RSOI and DSOI that leads to the anisotropic
spin-splitting subbands. The different Fermi wavevectors for the
spin-up and spindown subbands results in the quantum interference
and the oscillations of the conductivity. With increasing the
strengths of the DSOI, the oscillations of the conductivity becomes
stronger. Notice that the conductivity does not depend on the
orientation of the wire for the RSOI alone [see the dotted blue line
in Fig. 2(a)], this feature can be understood from Eq. (1) that does
not contain the angle $\theta $. Surprisingly, when the Coulomb
interaction is stronger, the anisotropy of the conductivity becomes
more obvious, which means that the Coulomb interaction enhance the
oscillation of the conductivity [see Fig. 2(b)].

The realistic quantum wire sample have finite lengths and are
connected adiabatically to the source and drain where
electron-electron interaction and SOI are negligible. Consider a
quantum wire of length $L$, and attached to two identical reservoirs
at its end points $x=0,L$. The charge conductivity is
\begin{eqnarray}
\sigma _{\rho }(x,x^{^{\prime }},\omega )&=&\frac{2e^{2}}{h}\Big(A_{1}e^{i%
\frac{\omega }{u_{1}}x}+A_{2}e^{-i\frac{\omega
}{u_{1}}x}\nonumber\\&+&A_{3}e^{i\frac{\omega
}{u_{2}}x}+A_{4}e^{-i\frac{\omega }{u_{2}}x}\Big), \label{myeq5}
\end{eqnarray}%
where $A_{i}(i=1,2,3,4)$ is function of $x^{^{\prime }}$ and $\omega
$ and can be deduced from the boundary conditions. From the above
equation, one can obtain the dc conductivity $\sigma _{\rho
}(0,L,0)=2e^{2}/h$. It means that the SOIs would not affect the dc
conductivity, since this behavior is caused by the contact
resistance at the ends of the quantum wire (see Eq. (55) in Ref.
[10]). And one actually could not observe any anisotropy of the
conductivity for a narrow and clean quantum wire. Consequently one
can not detect and distinguish the strengths of the RSOI and DSOI.
However, the ac conductivity is very different from the dc
conductivity. By tuning the strengths of the SOIs, the ac
conductivity can still exhibit the anisotropy, which can be clearly
seen from Fig. 3. From the nodes of the beating pattern of the ac
conductivity, one can determine the values of $u_{1}$ and $u_{2}$.
Then we can further calculate the value of $\delta v$ from Eq. (4). When $%
\theta =0$, there is $\delta v=\alpha $, which means that we can
obtain directly the value of $\alpha $ from the positions of the
nodes. And $\delta v=\sqrt{\alpha ^{2}+\beta ^{2}/4}$ corresponds to
crystallographic direction $\theta =\pi /2$. From the positions of
the nodes, we finally can obtain the values of both $\alpha $ and
$\beta $.
\begin{figure}
\center
\includegraphics[width=1\columnwidth]{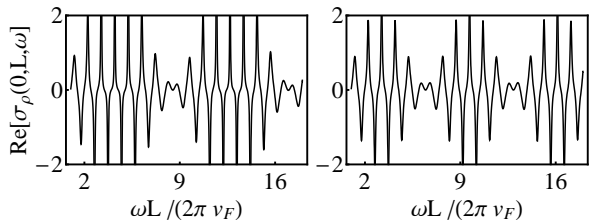} %
\caption{The ac conductivity (in units of $e^{2}/h$) as a function of $%
\protect\omega L/(2\protect\pi v_{F})$ for fixed $\protect\alpha $=0.02, $%
\protect\beta $=0.1, $g$=0.1 and different angles $\protect\theta $, Left: $%
\protect\theta $=$0$, Right: $\protect\theta $=$\protect\pi /2$,
respectively.}\label{fig:fig3}
\end{figure}

In conclusion, we investigate theoretically the charge transport
property of semiconductor quantum wires oriented in different
crystallographic directions in the presence of both the RSOI and
DSOI. We find that the Coulomb interaction can enhance the
anisotropy of the dc conductivity in an infinite long quantum wire.
In contrary to the Fermi liquid, the anisotropy of the dc
conductivity in LL is smeared out in a quantum wire with finite
length attached to two reservoirs, which make it impossible to
detect the relative strength of RSOI and DSOI. Instead, the ac
conductivity exhibits anisotropic behavior and interesting beating
patterns with increasing the radiation frequency. From the node
positions of the beating pattern of the ac conductivity, one can
detect and distinguish the strengths of the RSOI and DSOI.

\begin{acknowledgements}
This work was supported by NSFC, and the KLLDQSQC (Hunan Normal
University), and the construct program of the key discipline in
Changsha University of science and technology.
\end{acknowledgements}

\end{document}